# The behaviour of the excess Ca II H & K and Hε emissions in chromospherically active binaries *

D. Montes, M.J. Fernández-Figueroa, M. Cornide, and E. De Castro

Departamento de Astrofísica, Facultad de Físicas, Universidad Complutense de Madrid, E-28040 Madrid, Spain
E-mail: dmg@ucmast.fis.ucm.es

**Abstract.** In this work we analyze the behaviour of the excess Ca II H & K and Hε emissions in a sample of 73 chromospherically active binary systems (RS CVn and BY Dra classes), of different activity levels and luminosity classes. This sample includes the 53 stars analyzed by Fernández-Figueroa et al. (1994) and the observations of 28 systems described by Montes et al. (1995c). By using the spectral subtraction technique (subtraction of a synthesized stellar spectrum constructed from reference stars of spectral type and luminosity class similar to those of the binary star components) we obtain the active-chromosphere contribution to the Ca II H & K lines in these 73 systems. We have determined the excess Ca II H & K emission equivalent widths and converted them into surface fluxes. The emissions arising from each component were obtained when it was possible to deblend both contributions.

We have found that the components of active binaries are generally stronger emitters than single active stars for a given effective temperature and rotation rate.

A slight decline of the excess Ca II H & K emissions towards longer rotation periods, $P_{\rm rot}$, and larger Rossby numbers, $R_0$, is found. When we use $R_0$ instead of $P_{\rm rot}$ the scatter is reduced and a saturation at $R_0 \approx 0.3$ is observed.

A good correlation between the excess Ca II K and Hε chromospheric emission fluxes has been found. The correlations obtained between the excess Ca II K emission and other activity indicators, (C IV in the transition region, and X-rays in the corona) indicate that the exponents of the power-law relations increase with the formation temperature of the spectral features.

**Key words:** stars: activity – stars: binaries: close – stars: chromospheres – stars: late-type – stars: rotation

## 1. Introduction

The chromospherically active binaries are detached binary systems with cool components characterized by strong chromospheric, transition region, and coronal activity. The high levels of activity observed in these systems are generally attributed to the presence of deep convection zones and the fast rotation that drive the dynamo mechanism. In these binaries, the stars are usually forced to rotate relatively rapidly from tidal interaction and typically have rotation periods that are synchronized with their orbital periods.

In this group of chromospherically active binaries we include the RS Canum Venaticorum (RS CVn) binary systems defined by Hall (1976) and the BY Draconis (BY Dra) stars defined by Bopp & Fekel (1977). The RS CVn systems have, at least a cool evolved component, whereas the components of the BY Dra binaries are main sequence stars (Fekel et al. 1986). General properties of these systems have been recently reviewed by Rodonò (1992); Guinan & Giménez (1993) and Barrado et al. (1994).

Strong emission cores in the Ca II H and K resonance lines, are the primary optical indicators of chromospheric activity, their source functions are collisionally controlled and hence are sensitive probes of electron density and temperature. By analogy with the Sun, these emissions are identified with enhanced chromospheric emission from plage-like regions and the chromospheric network, which show magnetic field strengths larger than those in surrounding zones. However, in the chromospherically active because it is necessary to take into account the contribution from both components to the observed spectrum. Some recent spectroscopic studies of Ca II H & K lines in RS CVn and BY Dra systems have been reported by Strassmeier et al. (1990), Fernández-Figueroa et al. (1994) (hereafter FFMCC) and references therein, Strassmeier (1994), and Montes et al. (1995c).

In spite of these difficulties to determine the level of activity, there are several reasons why binary systems play an important role in the study of magnetic activity and dynamo models. On one hand, the membership of an active star in a binary system, permits to determine its physical properties, and on the other hand tidal forces in these systems can cause the star to spin up rapidly, maintaining a high level of activity throughout most of its lifetime. Because of this, these systems become laboratories for studying high levels of activity and for testing stellar dynamo models and the importance of differential rotation to the dynamo.

In this paper we report an analysis of the Ca II H & K lines as a chromospheric activity indicator in a sample of 73 northern active binary systems selected from "A Catalog of Chromospherically Active Binary Stars (second edition)" (Strassmeier et al. 1993, hereafter CABS). This sample includes the 53 stars analysed by Fernández-Figueroa et al. (1994) and the new observations of 28 systems described by Montes et al. (1995c). By using both the reconstruction of the absorption line profile and the spectral subtraction technique, we have determined the excess Ca II H & K emissions for each system.

We have computed absolute chromospheric fluxes in Ca II H & K in order to estimate the contribution of these lines to the total energy emitted from the stellar chromospheres. For this sample of chromospherically active binaries we studied the dependence of the excess Ca II H & K emissions on stellar parameters such as the effective temperature and the rotational period. We also studied the relation with the excess H$\epsilon$ emission, determined in our spectra, and other transition region, and coronal activity indicators.

In § 2 we give the details of our observations and data reduction. In § 3 we discuss the dependence of the excess Ca II H & K emissions on effective temperature, rotation, and other activity indicators, and in § 4 we give the conclusions. For the plots of the Ca II H & K spectra and the description of the individual results of the star sample the reader if referred to FFMCC and Montes et al. (1995c).

## 2. Observations

The spectroscopic observations that we analyse in this paper are the result of a program devoted to the study of spectroscopic properties of chromospherically active binary stars, in the region of the Ca II H & K lines which started in 1985. The high resolution spectra of the Ca II H m Telescope at the German Spanish Astronomical Observatory (CAHA) in Calar Alto (Almería, Spain), using a Coudé spectrograph with the f/3 camera, and the Isaac Newton Telescope (INT) at the Observatorio del Roque de Los Muchachos (La Palma, Spain), using the Intermediate Dispersion Spectrograph (IDS). The details of the observations and data reduction for the different observational seasons from 1985 to 1993 can be found in FFMCC and Montes et al. (1995c).

### 2.1. Measured parameters

The excess Ca II H & K and H$\epsilon$ emissions equivalent widths (EW) have been measured by reconstruction of the absorption line profile (described by FFMCC) and using the spectral subtraction technique (explained by Montes et al. (1994, 1995a,c)). The Ca II H & K surface flux, $F_S$(Ca II H;K) has been obtained using the linear relationship between the absolute surface flux at 3950 Å (in erg cm$^{-2}$ s$^{-1}$ Å$^{-1}$) and the colour index (V-R) by Pasquini et al. (1988).

Tables 1, 2 and 3 give the Ca II H & K and H$\epsilon$ line parameters, measured in the observed and subtracted spectra of the sample of active stars, for the groups 1, 2, and 3 respectively. In this tables we include the stars analysed by FFMCC but now measured using the spectral subtraction technique and the stars analysed, using this technique, by Montes et al. (1995c). Column (3) gives the orbital phase ($\varphi$) for each measured spectrum, and in column (4), H and C mean emission belonging to hot and cool component respectively, and T means that at these phases the spectral features can not be deblended. Column (5) gives the weights for the hot and cool component ($S_H$ and $S_C$). In columns (6) and (7) we list the EW for the K and H lines, obtained by reconstruction of the absorption line profile, and in columns (8) to (10) we give the EW for the K, H and H$\epsilon$ lines, measured in the subtracted spectrum. The EW(H;K;H$\epsilon$) corrected for the contribution of the components to the total continuum are given in columns (11) to (13) and the Ca II H & K and H$\epsilon$ surface fluxes are given in columns (14) to (16).

The Ca II H & K and H$\epsilon$ line parameters for the reference stars, the active single stars and the active components of visual binaries can be found in Table 4.

## 3. Discussion

The excess Ca II H & K emissions in a sample of 73 chromospherically active binaries have been studied.

The sample stars have been divided in three groups according to the assigned luminosity class of the active component. We can summarize the results obtained for these groups as follows:

Group 1 (all components are dwarfs) contains 21 binaries, 12 of which show double-emission of the Ca II H

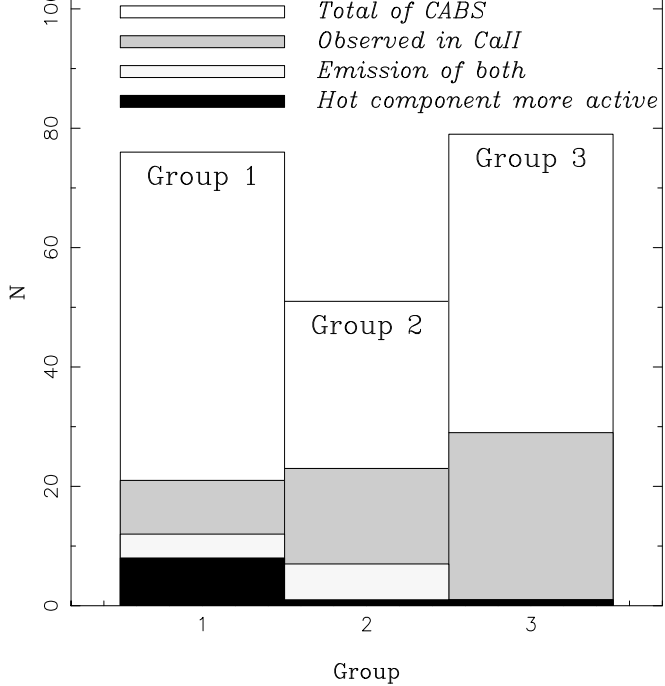

**Fig. 1.** Different behaviour of the Ca II H & K emissions in the groups 1, 2 and 3.

& K lines and in 5 the emission belongs to the hot component. When both components have the same or very similar spectral type the observed emissions are also very similar (YY Gem, BF Lyn, AS Dra, HD 108102, TZ CrB, KZ And). However, when the components of the system have different spectral types, the hot component tend to be the more active star of the system (DH Leo, V772 Her, BY Dra, KT Peg) or even the only active component as is the case of MS Ser, V815 Her, V775 Her, and V478 Lyr. Eleven systems show the H$\epsilon$ line in emission.

Group 2 (one active subgiant) contains 23 systems. In most cases the cool component is responsible for the emission. Seven systems exhibit Ca II H & K emissions coming from both components, although the cool component tend to be the more active star of the system. Twelve systems present the H$\epsilon$ line in emission.

Group 3 contains 29 systems. They are mainly single-lined binaries with single-emission of the Ca II H & K lines. In other systems the hot component is an A-type star (as is the case of RZ Eri, 93 Leo, $\epsilon$ UMi, HR 7428) or a white dwarf star (AY Cet, DR Dra). Two systems in this group (V1817 Cyg, and V1764 Cyg) present a clear self-absorption with blue asymmetry. Only in one system (RZ Cnc) the emission belongs to the hot component. In this group only eight systems present the H$\epsilon$ line in emission.

Fig. 1 is an histogram showing the number of systems observed by us in the Ca II H & K lines region, included in each group, with respect to the total number of systems in CABS and the different behaviour of the Ca II H & K component originating the emission.

In what follows we study the behaviour of the excess Ca II H & K emissions as a function of stellar parameters and the relation with other activity indicators. When several observations of the same star are available, a mean value is used in the analysis. In the following analysis we have used the EW(H;K) corrected for the contribution of the components to the total continuum. All linear regressions used in the discussion have been determined with the bisector method described by Isobe et al. (1990).

### 3.1. Dependence of the Ca II H & K emissions on effective temperature

In Fig. 2 (upper panel) we have plotted the logarithm of the excess Ca II K emission EW against the effective temperature ($T_{\text{eff}}$), which was derived from the spectral type-temperature relation from Schmidt-Kaler (1982) in Landolt-Börnstein. In this figure a slight trend of increasing EW(Ca II K) with cooler effective temperature can be seen, since it is possible to find very different activity levels for each temperature range. The active binary components are generally stronger emitters than single active stars (plotted in this figure with large open circles) of similar effective temperature.

When plotting the Ca II K surface flux, $F_S$(Ca II K), instead of EW(Ca II K) (Fig. 2, lower panel) a slight trend of decreasing $F_S$(Ca II K) with lower effective temperature can be seen. The $F_S$(Ca II K) are found to lie in the range $10^5$ - $10^8$ erg cm$^{-2}$ s$^{-1}$. In this figure dashed lines indicate constant $F_S$(Ca II K)/$F_{\text{bol}}$ ratios. The observed spread in both figures can be due to the fact that the Ca II K indicator can change with orbital phase and/or larger time scales (cycles) and to the dependence of the Ca II K emission on the rotational velocity in the sense that faster rotators show larger excess Ca II K emission for a given $T_{\text{eff}}$. In Fig. 2 (lower panel) the size of the symbols is inversely proportional to the rotational period.

### 3.2. Dependence of the Ca II H & K emissions on rotation

The dynamo mechanism involves interaction between rotation, differential rotation and convection, and provides the theoretical ground on the role played by rotation in determining the stellar magnetic activity.

In Fig. 3 we show the dependence of activity on surface rotation for the chromospherically active binaries and for the single active stars. In this figure we have plotted the excess Ca II K emission EW, log EW(Ca II K) (upper panel), and the excess Ca II K surface flux, log $F_S$(Ca II K) (lower panel), versus rotational period, log $P_{\text{rot}}$ (in days). The rotational periods ($P_{\text{rot}}$) are taken from photometric observations ($P_{\text{phot}}$). In the case of short-period binaries without a photometric period determination we have

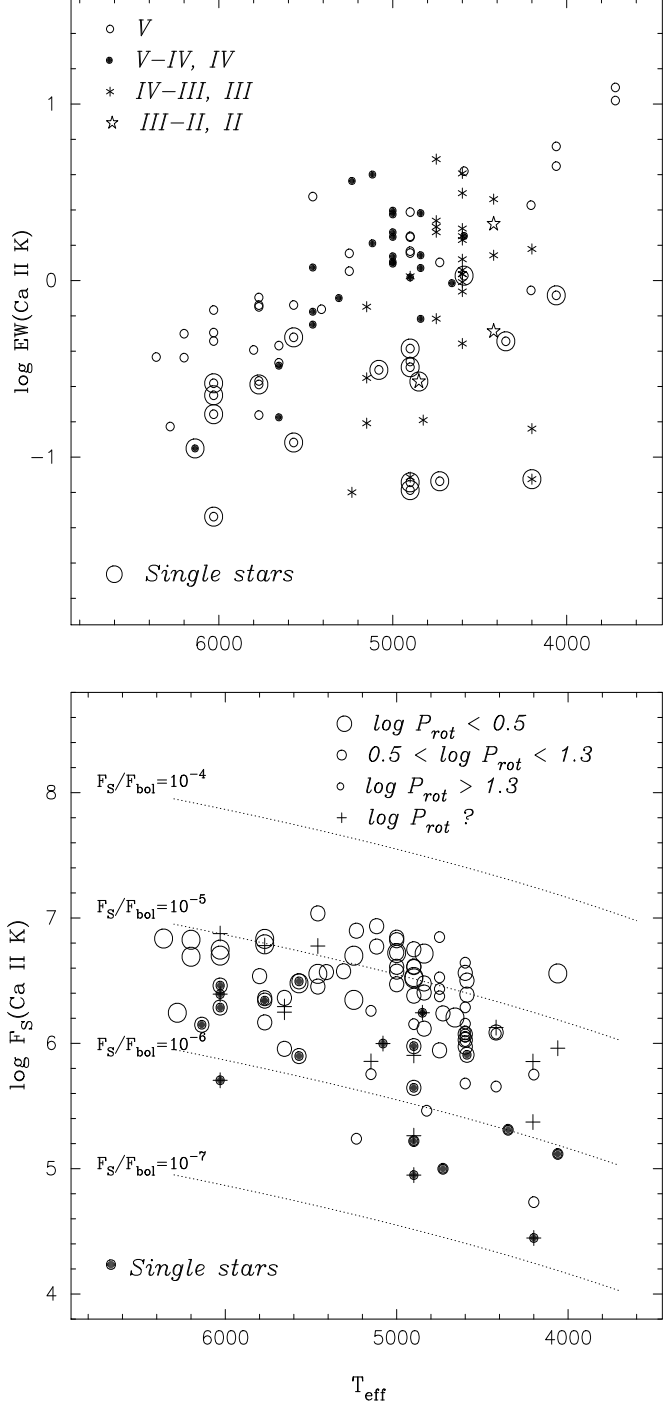

**Fig. 2.** Excess Ca II K emission EW, EW(Ca II K), (upper panel) and logarithm of Ca II K surface flux, $F_S$(Ca II K), (lower panel) plotted against effective temperature, $T_{eff}$. In the upper panel different symbols are used to represent the stars with different luminosity class. In the lower panel the size of the symbols is inversely proportional to the rotational period ($P_{rot}$). The dashed lines indicate constant $F_S$(Ca II K)/$F_{bol}$ ratios ($10^{-7}$ to $10^{-4}$)

periods ($P_{orb}$).

In both panels of Fig. 3 we can see that in spite of the large scatter in the excess Ca II K emission for each rotational period, a slight decline toward longer rotational periods is present. The scatter is lower when we plot surface flux instead of EW and the tendency of decreasing flux with increasing period is much more clear for the single active stars (plotted in this figure with large open circles). Again we see that the components of active binaries are generally stronger emitters than single active stars for a given rotational rate.

When binning our sample into narrow intervals of effective temperature (as indicated in Fig. 3 (upper panel) with different symbols), we can see that the lower effective temperature stars seem to have larger excess Ca II K EW for the same rotational period. This trend is mainly determined by the decrease of the continuum at the H & K line region with the decrease of the temperature.

Strassmeier et al. (1990), using single and binary stars, have found clear evidence in the sense that evolved stars have larger Ca II K emission fluxes than main-sequence stars at the same rotational period. We did not find this segregation in luminosity classes in our sample of binaries, as can be seen in Fig. 3 (lower panel) where we represent with different symbols stars with different luminosity class. In general dwarfs show shorter rotational periods than giants and subgiants, and there are neither dwarfs with large rotational periods nor giants with short rotational periods. Obviously, this is largely determined by selection effects in our sample. For instance, the dwarfs chromospherically active binaries are systems with nearly circular orbits and synchronization between orbital and rotation period, beeing their evolutionary status close to the TAMS (Terminal Age Main Sequence) or evolving off the main sequence (Barrado et al. 1994). A similar behaviour was seen in the case of the dependence of the H$\alpha$ emission on rotation in chromospherically active binaries (Montes et al. 1995a).

If we include in the study the single active stars we can see that there are single dwarfs with large rotational periods that lie below the giant binaries of the same rotational period. However, we think that one must be very careful in deriving conclusions based on comparisons between active binaries and single stars, since the dynamo mechanism should be affected by the internal rotation, that is essentially determined by the rotational history, which is not the same for single and binary stars, and also by the fact that very different physical processes in general would be found in single and binary stars.

In Fig. 3 we can also see that for a given rotation rate the range of Ca II K emission fluxes observed seem to be larger for the giant stars. The reason for that might be the pronounced differences between giants and dwarfs in the evolutionary, angular momentum loss and internal angular momentum redistribution time scales. Strassmeier et tified the relationship between rotation and activity for evolved stars, found a large range of emission-line fluxes for a given rotation rate and that the flux from the cooler stars depends stronger upon rotation than the flux from the hotter stars.

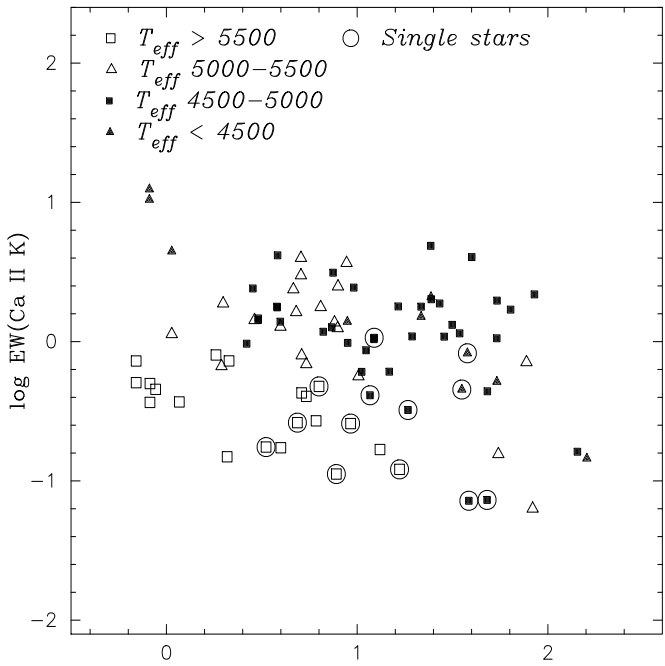

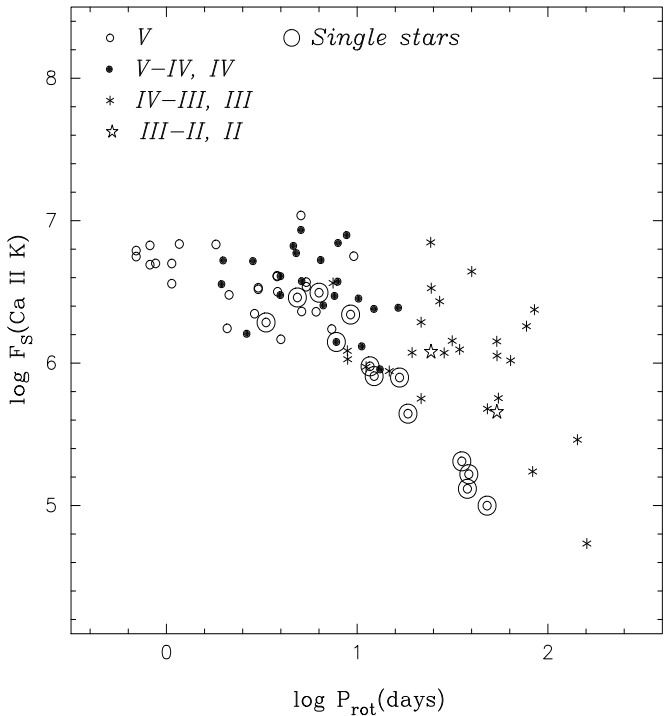

**Fig. 3.** log EW(Ca II K), (upper panel) and log $F_S$(Ca II K), (lower panel) plotted against logarithm of rotation period, $P_{rot}$. In the upper panel different symbols are used to represent the stars included in four temperature intervals. In the lower panel different symbols are used to represent the stars with different luminosity class.

On the other hand the rotation-activity relation could be determined by the strong dependence of radius on period found in this systems (Dempsey et al 1993), that results primarily from the fact that these systems are synchronous binaries covering a narrow range in effective temperature.

### 3.2.1. Dependence on the Rossby number, $R_0$

The rotation and convection can be parameterized by the Rossby number, $R_0$, which is defined as the ratio of the rotation period, $P_{rot}$, to the turnover time of a convective cell at the base of the convection zone, $\tau$, ($R_0 = P_{rot}/\tau$). This is crudely related to dynamo activity levels according to Durney and Latour (1978) and is an useful index for quantifying the efficiency of the interaction of the stellar rotation and convection to produce magnetic fields. Noyes et al. (1984) showed that the use of $R_0$ instead of $P_{rot}$ reduced the scatter in correlations with the Ca II H & K emission in main sequence stars. However, not all observers accept $R_0$ as a better parameter than $P_{rot}$ (Basri 1986; Young et al. 1989; Strassmeier et al. 1990; Rutten 1987).

The Rossby number for the stars in our sample has been determined as follows. For dwarf stars we have adopted the semi-empirical turnover time values given by Noyes et al. (1984) as a function of B-V, the values obtained in this way differ only slightly from the theoretical ones given by Gilman (1980) for the ratio of mixing length to the density scale height, $\alpha = 1.9$. For subgiant and giant stars we have used the turnover times given by Basri (1987) as a function of effective temperature which was adopted from Gilliland (1985) after correction of the effective temperature scale.

In Fig. 4 (upper panel) we have plotted the logarithm of $F_S$(Ca II K) versus the logarithm of the Rossby number, $R_0$. As shown, there is a decline toward larger Rossby numbers for the more slowly rotating stars, with a scatter lower than in Fig. 3. For the more rapidly rotating active stars we observe that the chromospheric Ca II K emission flux saturate at Rossby number $R_0 \approx 0.3$ ($\log R_0 \approx -0.5$). A similar saturation limit has been found, using chromospheric UV and optical line flux indicators (Vilhu 1984), soft X-ray observations of the transition region (Vilhu & Rucinski 1983) and using unpolarized Zeeman broadened line profiles (Saar 1991). This saturation limit has been interpreted by Vilhu (1984) as the total filling of the surface with emitting structures and active regions. However O'Dell et al. (1995), studying the maximum amplitude of photometric variability in young solar-type stars esti-a factor 6 to 10 times lower than the inferred from chromospheric indicators.

When we plot $R(Ca\ \textsc{ii}\ H+K) = F_S(Ca\ \textsc{ii}\ K)/\sigma\ T_{\rm eff}^4$, instead of $F_S(Ca\ \textsc{ii}\ K)$ (Fig. 4, lower panel) the scatter is lower for the single active stars, however, for the binary systems the scatter is nearly the same. In this figure we have also plotted the relation between $R(Ca\ \textsc{ii}\ H+K)$ and $R_0$ given from Noyes et al. (1984) (solid-line), Soderblom (1985) (dotted-dashed line) and Montesinos & Jordan (1993) (dashed line). As we can see, in this figure, only the single active stars seem to fit well these relationships, while the binary systems components, dwarfs and giants, exhibit $R(Ca\ \textsc{ii}\ H+K)$ values well above the above mentioned relationships. Zwaan (1991) also found these high levels of activity in the RS CVn systems in comparison with single stars with comparable spectral types and rotation periods, and suggests that this is somehow produced by enhancements in the dynamo efficiency by tidal interactions between the components of the binary system.

### 3.3. Ca II K – Hε relation

We study in our sample the behaviour of the Hε emission as an alternative activity indicator and the relation of this line with the excess Ca II K emission. The excess Hε emission EW has been determined using the same spectral subtraction technique as in the case of Ca II H & K lines. In some cases Gaussian fits have been performed in order to separate the Hε line from the Ca II H line. The Hε surface flux, $F_S(H\epsilon)$, has been determined with the calibration of Pasquini et al. (1988) for the Ca II H & K lines, since the Hε line appears in the same spectral region.

Fig. 5 shows that the Hε emission surface flux is well correlated with the Ca II K emission surface flux. We have obtained the following linear regression line:

$$\log F_S(H\epsilon) = (1.17 \pm 0.13) \log F_S(Ca\ \textsc{ii}\ K) \\ - (1.70 \pm 0.84) \quad (r = 0.95). \quad (1)$$

Due to the simultaneity of the observations in Hε and Ca II K lines, the observed scatter is lower that in the case of the relation between Hε and Hα lines found by Montes et al. (1995a).

Furthermore, in our study of the excess Hα emission (Montes et al. 1995a) we have found that all stars of our sample with Hα emission above the continuum have the Hε line in emission and, for the same $T_{\rm eff}$ the stars with Hε in emission have larger excess Hα emission and also have larger excess Ca II K emission. So we can conclude that the Hε emission line is an alternative activity indicator.

### 3.4. Ca II K and other activity indicators

Flux-flux relationships between activity indicators originated in different temperatures regimes of the stellar outer atmospheres (chromosphere, transition region and corona) have been extensively studied in the literature for different kinds of stars (Ayres et al. 1981, 1995; Oranje et al. 1982; Marilli & Catalano 1984; Basri 1987; Montesinos & Jordan 1988; Rutten et al. 1991, Schrijver et al. 1992). These studies have shown that, as the energy loss from the chromosphere increases, the total energy loss from the overlying transition region and corona increases more rapidly with increasing plasma temperature, and that these relationships are generally non-linear. These nonlinear power laws are difficult to reconcile with a straightforward "solar analog" that explains enhanced emissions of stars through increased surface coverage by some fundamental quantum of magnetic activity.

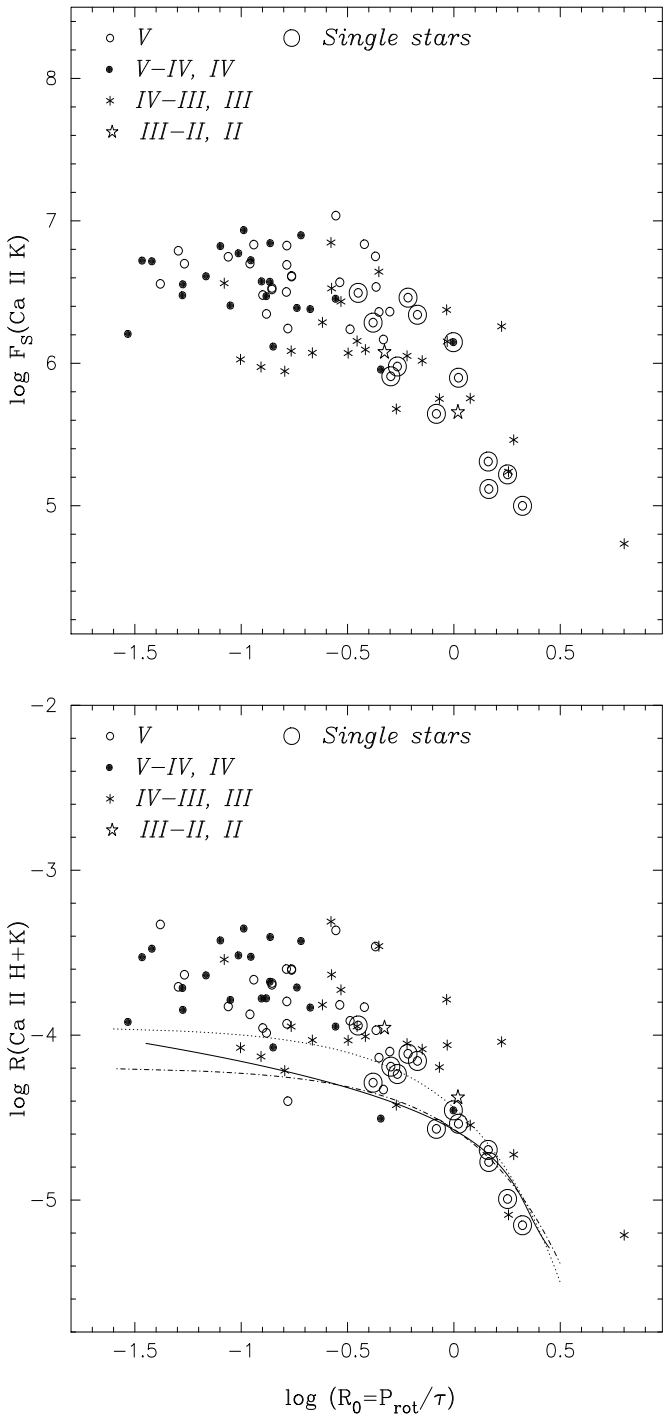

**Fig. 4.** log $F_S$(Ca II K) (upper panel) and log R(Ca II H+K) (lower panel), plotted versus the logarithm of Rossby number, $R_0 = P_{\rm rot}/\tau$. Different symbols are used to represent the stars with different luminosity class. In the lower panel we have plotted with different type of lines the relation between R(Ca II H+K) and $R_0$ given from different authors.

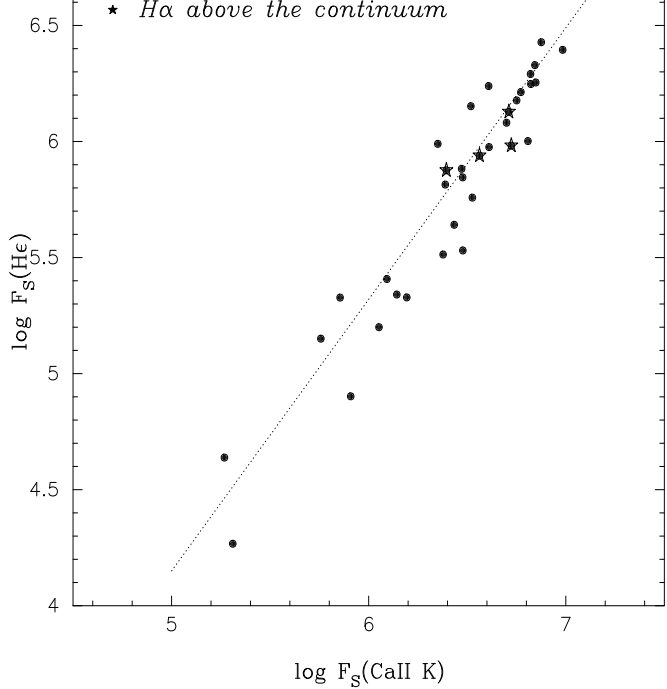
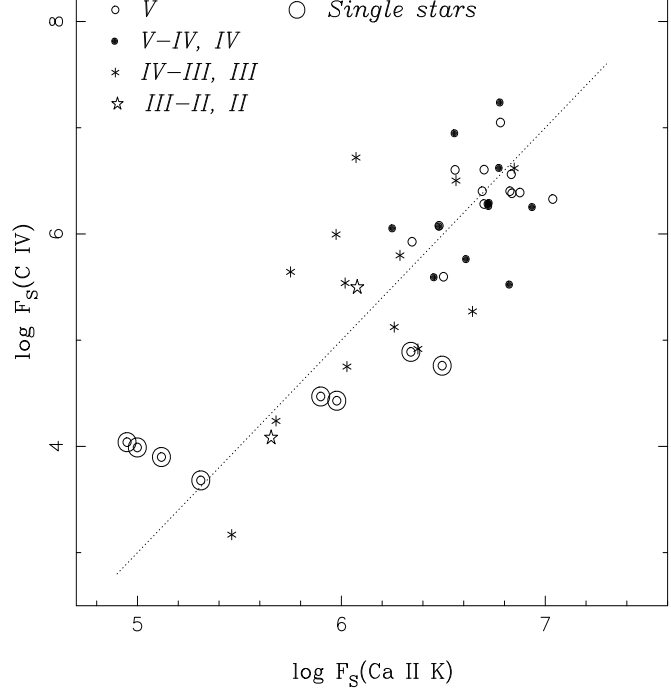

**Fig. 5.** log $F_S(H\epsilon)$ versus log $F_S(Ca\,II\,K)$. The dotted line corresponds to the best linear fit.

**Fig. 6.** log $F_S(C\,IV)$ versus log $F_S(Ca\,II\,K)$. Different symbols are used to represent the stars with different luminosity class. The dotted line corresponds to the best linear fit.

In the following we use our Ca II K line flux data to study the flux-flux relationships between this chromospheric activity indicator and the transition region C IV line, and the coronal X-ray emission in our sample of chromospherically active binary systems. When both components are active and the C IV and X-ray emission fluxes refer to both components, we also take the Ca II K cumulated flux from both components in the plots.

### 3.4.1. Ca II K – C IV relation

We study here the relation between the chromospheric excess Ca II K emission and the transition region C IV lines ($\lambda$ 1548, 1550 Å) in our sample of binary systems.

We obtained the C IV line fluxes from the low resolution IUE (International Ultraviolet Explorer) spectra stored in the IUE uniform low dispersion archive (ULDA) when available, for the remaining systems they were taken from Basri et al. (1985) and Fox et al. (1994). The absolute fluxes at the stellar surface, $F_S(C\,IV)$, have been determined using the stellar data from CABS. For single active stars we have taken the C IV line fluxes given by Basri (1987) and Ayres et al. (1995). We have plotted in Fig. 6 log $F_S(C\,IV)$ against log $F_S(Ca\,II\,K)$ and the following linear regression for the chromospheric active binary systems has been found:

$$\log F_S(C\,IV) = (2.00 \pm 0.21) \log F_S(Ca\,II\,K) \\ - (7.00 \pm 1.41) \quad (r = 0.71). \quad (2)$$

The single active stars seem to present a different relation. In this figure different symbols are used to represent stars with different luminosity class. As shown, the dwarfs and subgiants binaries present the larger C IV lines fluxes. The scatter in this figure can be partially due to the non simultaneity of the observations and to the temporal variations of both emission line fluxes.

### 3.4.2. Ca II K – X-rays relation

Ca II K emission flux has also been found to be well correlated with the coronal X-ray emission flux. The power law relation between these two fluxes has been discussed by a number of authors. Early data gave the exponent ≈ 3 (Ayres et al. 1981). Later Schrijver (1987) and Schrijver et al. (1992) found it to be ≈ 1.5 whereas Rutten et al. (1991) determined a value between 2.2 and 2.6. Recently Stępień (1994) have obtained a value of 3.4.

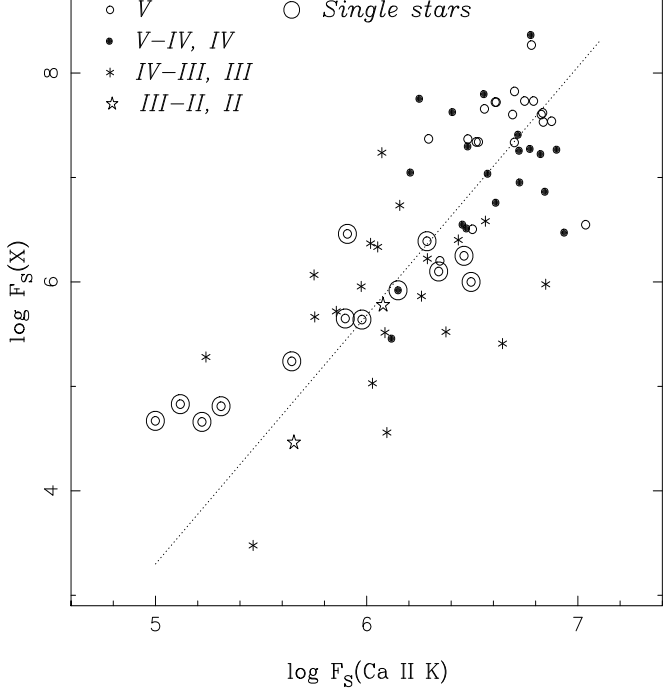
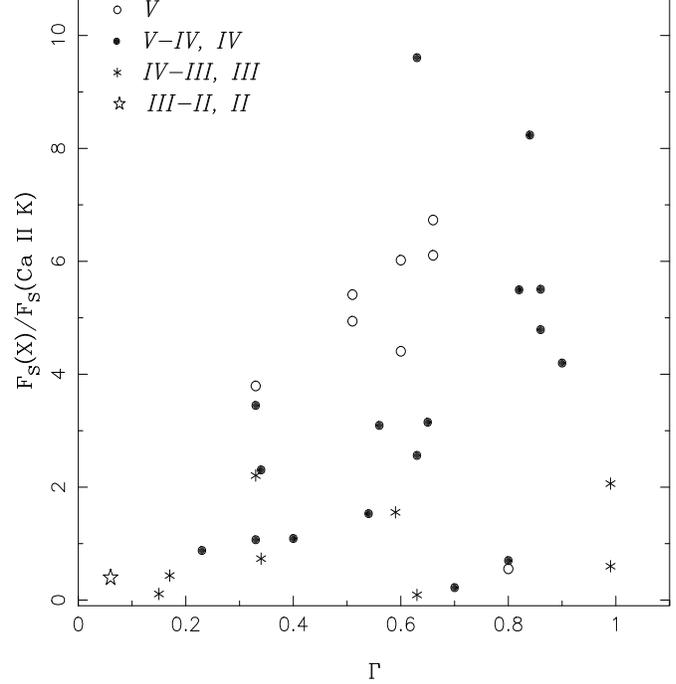

**Fig. 7.** log F$_S$(X) versus log F$_S$(Ca II K). Different symbols are used to represent the stars with different luminosity class. The dotted line corresponds to the best linear fit.

**Fig. 8.** F$_S$(X)/F$_S$(Ca II K) versus Γ. Different symbols represent stars with different luminosity class.

We look for a relation between these two activity indicators in the stars of our sample.

The X-ray data used are from ROSAT (Röntgensatellite) PSPC (Position Sensitive Proportional Counter) all-sky survey. For binary systems we have taken the X-ray luminosities from Dempsey et al. (1993, 1994) although, in some cases, the values have been modified taking into account the new distances tabulated in CABS. The absolute flux at the stellar surface, F$_S$(X), has been determined using the stellar radii from CABS. For single active stars we have taken the X-ray surface fluxes given by Hempelmann et al. (1995).

We plot, in Fig. 7, log F$_S$(X) against log F$_S$(Ca II K) for the stars of our sample, different symbols are used to represent stars with different luminosity class. As shown, the dwarfs and subgiants present the larger X-ray emission lines fluxes. A correlation between both activity indicators is found in the chromospheric active binary systems. The dotted line, in this figure represents the following linear regression:

$$\log F_S(X) = (2.38 \pm 0.26) \log F_S(\text{Ca II K}) \\ - (8.60 \pm 1.66) \quad (r = 0.77). \quad (3)$$

This correlation can be written as a power law relation with an exponent of 2.38 which is within the range of values given for this exponent by the above mentioned authors. For the single active stars the exponent of the power law relation seems to be lower. The scatter present in this figure, larger for the binary systems, can be partially due to the non simultaneity of the observations and the temporal variations of both activity indicators in these systems.

The X-ray emission from RS CVn systems is generally believed to arise from coronal loop structures. However, in semi-detached and high Roche lobe filling fraction systems it could be a relation between mass transfer and X-ray emission as has been recently suggested by Welty & Ramsey (1995).

In Fig. 8, we have plotted the ratio between F$_S$(X) and F$_S$(Ca II K) versus the Roche lobe filling fraction, Γ = R$_*$/R$_\text{Roche}$, for the systems with known stellar and orbital parameters. In this figure we can see that the high - Γ systems tend to have large ratios between the coronal X-ray flux and the chromospheric Ca II K flux, suggesting that mass transfer could be responsible for significant X-ray emission in these systems, similarly to the result of Welty & Ramsey (1995) using the X-ray luminosities.

We have determined excess Ca II H & K emissions EW for 73 chromospherically active binaries using the spectral subtraction technique. Whenever possible the emission arising from each component was obtained. In groups 2 and 3 of the sample selected the bulk of the emission comes from the coolest and most evolved component, or from both components, if they are similar. However, in group 1 the hot component tend to be the more active star of the system.

The behaviour of the excess Ca II H & K emission as a function of effective temperature and rotation period is very similar to that found in the H$\alpha$ line by Montes et al. (1995a). A slight decline towards longer rotation periods, $P_{rot}$, and larger Rossby numbers, $R_0$, is found, The scatter is lower when $R_0$ is used instead of $P_{rot}$. The active binary components are generally stronger emitters than single active stars for a given rotation rate. For the more rapidly rotating active stars we observe that the chromospheric Ca II K emission fluxes saturate at Rossby number $R_0 \approx 0.3$. On the other hand, the components of active binary systems, dwarfs and giants, exhibit R(Ca II H+K) values well above those expected from the relationships between R(Ca II H+K) and $R_0$ found by other authors.

We have compared the derived Ca II K fluxes with those measured in the H$\epsilon$ line finding that a good correlation exists between these chromospheric activity indicators. The H$\epsilon$ emission line appears to be an alternative activity indicator.

Flux-flux relationships between activity indicators originated in different temperature regimes of stellar outer atmospheres (C IV in the transition region and X-rays in the corona) have been also analysed. The correlations obtained between the different activity indicators and the excess Ca II K emission can be written as flux-to-flux power-law relations as follows:

$$F_S(H\epsilon) \propto F_S(Ca\,II\,K)^{1.17}. \quad (4)$$

$$F_S(C\,IV) \propto F_S(Ca\,II\,K)^{2.00}. \quad (5)$$

$$F_S(X) \propto F_S(Ca\,II\,K)^{2.38}. \quad (6)$$

As it can be see, the exponents of these power-law relations increase with the temperature of the regions where the spectral features originate. A similar result was obtained by Rutten et al. (1991) in their study relative to the Ca II H+K flux and by Montes et al. (1995a) relative to the excess H$\alpha$ emission.

Our survey provides a large body of new high quality data on Ca II H & K emissions in chromospherically active binaries. However, a large number of simultaneous different orbital phases are necessary for a better understanding of the behaviour of these activity indicators and in order to analyse the flux-flux diagrams and minimize the effects of time-dependent processes, such as activity cycles, rotational modulation, active region evolution, etc. that contribute to the scatter in the flux-flux diagrams.

*Acknowledgements.* We thank F. Martín for the C IV flux measurements in IUE spectra retrieved from ULDA. We would also like to thank Dr. B. Montesinos for the very careful reading of the manuscript and the referee S. Catalano for suggesting several improvements and clarifications. This work has been supported by the Universidad Complutense de Madrid and the Spanish Dirección General de Investigación Científica y Técnica (DGICYT) under grant PB94-0263.

## References

Ayres T.R., Marstad N.C., Linsky J.L. 1981, ApJ 247, 545
Ayres T.R., Fleming T.A., Simon T., et al. 1995, ApJS 96, 223
Barrado D., Fernández-Figueroa M.J., Montesinos B., De Castro E. 1994, A&A 290, 137
Basri G. 1986, *Cool Stars, Stellar Systems, and the Sun, Fourth Cambridge Workshop*, M. Zeilik, D.M. Gibson (eds.), Springer-Verlag, p. 184
Basri G. 1987, ApJ 316, 377
Basri G., Laurent R., Walter F. M. 1985, ApJ 298, 761
Bopp B.W., Fekel F.C. 1977, AJ 82, 490
Dempsey R.C., Linsky J.L., Fleming T.A., Schmitt J.H.M.M. 1993, ApJS 86, 599
Dempsey R.C., Linsky J.L., Fleming T.A., Schmitt J.H.M.M. 1994, in: Cool Stars, Stellar Systems, and the Sun, Eighth Cambridge Workshop, J.P. Caillault (ed.), ASP Conference Series 64, 74
Durney B.R., Latour J. 1978, *Geophys. Ap. Fluid Dyn.* 9, 241
Fekel F.C., Moffett T.J., Henry G.W. 1986, ApJS 60, 551
Fernández-Figueroa M.J., Montes D., De Castro E., Cornide M. 1994, ApJS 90, 433 (FFMCC)
Fox D.C., Linsky J.L., Veale A. et al. 1994, A&A 284, 91
Gilman P., 1980, in IAU Colloq. 51, Stellar Turbulence, D. Gray and J. Linsky (ed) (Springer), p.19
Gilliland R.L. 1985, ApJ 300, 339
Guinan E.F., Giménez A. 1993, in The Realm of Interacting Binary Stars, J. Shade et al. (eds.), Kluwer Academic Publishers, p. 51
Hall D.S. 1976, in IAU Symp., Multiple Periodic Variable Stars, ed. W. S. Fitch (Dordrecht: Reidel), p. 287
1990, ApJ 358, 610
Hempelmann A., Schmitt J.H.M.M., Schultz M., Rüdiger G., Stępień K. 1995, A&A 294, 515
Isobe T., Feigelson E.E., Akritas M.G., Babu G.J. 1990, ApJ 364, 104
Marilli E., Catalano S. 1984, A&A 133, 57
Montes D., De Castro E., Cornide M., Fernández-Figueroa M.J. 1994, in: Cool Stars, Stellar Systems, and the Sun, Eighth Cambridge Workshop, J.P. Caillault (ed.), ASP Conference Series 64, 444
Montes D., Fernández-Figueroa M.J., De Castro E., Cornide M. 1995a, A&A 294, 165

M. 1995b, A&AS 109, 135

Montes D., De Castro E., Fernández-Figueroa M.J., Cornide M. 1995c, A&AS 114, 287

Montesinos B., Jordan C. 1988, *A Decade of UV Astronomy with the IUE Satellite*, ESA SP-281, 1, 238

Montesinos B., Jordan C. 1993, MNRAS 264, 900

Noyes R.W., Hartmann L.W., Baliunas S.L., Duncan D.K., Vaughan A.H. 1984, ApJ 279, 763

O'Dell M.A., Panagi P., Hendry M.A., Collier Cameron A. 1995, A&A 294, 715

Oranje B.J., Zwaan C., Middlekoop F. 1982, A&A 110, 30

Pasquini L., Pallavicini R., Pakull M. 1988, A&A 191, 253

Rodonò M. 1992, *Evolutionary Processes in Interacting Binary Stars*, Y. Kondo et al. (eds.), p. 71

Rutten R.G.M. 1987, A&A 177, 131

Rutten R.G.M., Schrijver C.J., Lemmens A.F.P., Zwaan C. 1991, A&A 252, 203

Saar S. 1991, *The Sun and Cool Stars: Activity, Magnetism, Dynamos*, I. Tuominen, D. Moss, G. Rüdiger (eds.), Springer-Verlag, p. 389

Schmidt-Kaler T. 1982, in Landolt-Börnstein, Vol. 2b, (eds.) K. Schaifers, H.H. Voigt (Heidelberg: Springer)

Schrijver C.J. 1987, A&A 172, 111

Schrijver C.J., Dobson A.K., Radick R.R. 1992, A&A 258, 432

Soderblom D.R. 1985, AJ 90, 2103

Stępień K. 1994, A&A 292, 191

Strassmeier K.G., 1994, A&AS 103, 413

Strassmeier K.G., Fekel F.C., Bopp B.W., Dempsey R.C., Henry G.W. 1990, ApJS 72, 191

Strassmeier K.G., Hall D.S., Fekel F.C., Scheck M. 1993, A&AS 100, 173 (CABS)

Vilhu O. 1984, A&A 133, 117

Vilhu O., Rucinski S.M. 1983, A&A 127, 5

Welty A.D., Ramsey L.W. 1995, AJ 109, 2187

Young A., Skumanich A., Stuffer J.R., Bopp B.W., Harlan E. 1989, ApJ 344, 427

Zwaan C. 1991, in Mechanisms of Chromospheric and Coronal Heating, P. Ulmschneider, E. Priest, R. Rosner (eds.), Springer-Verlag Berlin p. 241

**Table 1.** Ca II H & K lines measures in the observed and subtracted spectrum (Group 1)

| Name | Date | $\varphi$ | E | $S_H/S_C$ | Reconstruction EW (K) | Reconstruction EW (H) | Spectral subtraction EW (K) | Spectral subtraction EW (H) | Spectral subtraction EW (H$\epsilon$) | Corrected EW (K) | Corrected EW (H) | Corrected EW (H$\epsilon$) | Absolute flux log F (K) | Absolute flux log F (H) | Absolute flux log F (H$\epsilon$) |
|---|---|---|---|---|---|---|---|---|---|---|---|---|---|---|---|
| 13 Cet | 23/11/86 | 0.17 | H | - | 0.041 | 0.020 | 0.149 | 0.154 | - | 0.15 | 0.15 | - | 6.24 | 6.26 | - |
| VY Ari | 16/12/92 | 0.17 | - | - | 1.643 | 1.412 | 1.789 | 1.614 | 0.477 | 1.79 | 1.61 | 0.48 | 6.40 | 6.34 | 5.81 |
| OU Gem | 04/03/93 | 0.47 | T | 0.74/0.26 | 0.774 | 0.715 | 0.938 | 0.920 | 0.232 | 1.27 | 1.24 | 0.31 | 6.24 | 6.23 | 5.63 |
| YY Gem | 04/03/93 | 0.44 | 1 | 0.50 | 5.049 | 4.901 | 5.232 | 5.378 | - | 10.5 | 10.8 | - | 5.51 | 5.52 | - |
|  |  |  | 2 | 0.50 | 6.033 | 7.613 | 6.209 | 7.104 | 3.199 | 12.4 | 14.2 | 6.40 | 5.59 | 5.65 | 5.30 |
| BF Lyn | 05/03/93 | 0.21 | H | 0.50 | 0.832 | 0.717 | 0.891 | 0.861 | - | 1.78 | 1.72 | - | 6.61 | 6.60 | - |
|  |  |  | C | 0.50 | 0.823 | 0.795 | 0.881 | 0.982 | 0.376 | 1.76 | 1.96 | 0.75 | 6.61 | 6.66 | 6.24 |
| DH Leo | 29/01/88 | 0.55 | H | 0.94 | 0.864 | 0.788 | 1.075 | 1.111 | 0.268 | 1.14 | 1.18 | 0.29 | 6.70 | 6.72 | 6.10 |
|  |  |  | C | 0.06 | 0.203 | 0.210 | 0.218 | 0.183 | - | 3.63 | 3.05 | - | 6.47 | 6.39 | - |
| " | 01/02/88 | 0.32 | H | 0.94 | 1.469 | 1.412 | 1.017 | 1.008 | 0.245 | 1.08 | 1.07 | 0.26 | 6.68 | 6.68 | 6.06 |
|  |  |  | C | 0.06 | 0.446 | 0.160 | 0.243 | 0.276 | - | 4.05 | 4.60 | - | 6.52 | 6.57 | - |
| " | 05/03/93 | 0.07 | H | 0.94 | 0.948 | 0.804 | 1.098 | 0.965 | 0.314 | 1.17 | 1.03 | 0.33 | 6.71 | 6.66 | 6.17 |
|  |  |  | C | 0.06 | 0.164 | 0.325 | 0.256 | 0.214 | - | 4.27 | 3.57 | - | 6.54 | 6.46 | - |
| " | 08/03/93 | 0.70 | H | 0.94 | 1.067 | 0.823 | 1.085 | 1.133 | 0.213 | 1.15 | 1.21 | 0.23 | 6.71 | 6.73 | 6.00 |
|  |  |  | C | 0.06 | 0.283 | 0.297 | 0.363 | 0.406 | - | 6.05 | 6.77 | - | 6.69 | 6.74 | - |
| " | 07/03/93 | 0.87 | H | 0.94 | 1.052 | 0.753 | 1.038 | 1.074 | 0.240 | 1.10 | 1.14 | 0.26 | 6.69 | 6.70 | 6.05 |
|  |  |  | C | 0.06 | 0.355 | 0.068 | 0.256 | 0.253 | - | 4.27 | 4.22 | - | 6.54 | 6.53 | - |
| $\xi$ UMa(B) | 28/01/88 | 0.26 | - | - | 0.145 | 0.125 | 0.193 | 0.169 | - | 0.19 | 0.17 | - | 6.21 | 6.16 | - |
| " | 29/01/88 | 0.49 | - | - | 0.127 | 0.122 | 0.173 | 0.159 | - | 0.17 | 0.16 | - | 6.17 | 6.13 | - |
| AS Dra | 07/03/93 | 0.49 | T | 0.66/0.34 | 0.311 | 0.281 | 0.444 | 0.383 | 0.208 |  |  |  |  |  |  |
| " | 09/03/93 | 0.85 | H | 0.66 | 0.174 | 0.205 | 0.267 | 0.272 | - | 0.41 | 0.41 | - | 6.54 | 6.54 | - |
|  |  |  | C | 0.34 | 0.152 | 0.137 | 0.234 | 0.235 | - | 0.69 | 0.69 | - | 6.57 | 6.57 | - |
| IL Com | 31/01/88 | * | 1 | 0.50 | 0.143 | 0.116 | 0.315 | 0.330 | - | 0.63 | 0.66 | - | 6.93 | 6.95 | - |
|  |  |  | 2 | 0.50 | 0.108 | 0.069 | 0.177 | 0.184 | - | 0.35 | 0.37 | - | 6.68 | 6.69 | - |
| " | 01/02/88 | * | 1 | 0.50 | 0.136 | 0.087 | 0.302 | 0.349 | - | 0.60 | 0.70 | - | 6.91 | 6.97 | - |
|  |  |  | 2 | 0.50 | 0.111 | 0.056 | 0.181 | 0.180 | - | 0.36 | 0.36 | - | 6.69 | 6.68 | - |
| " | 05/03/93 | * | 1 | 0.50 | 0.115 | 0.072 | 0.183 | 0.154 | - | 0.37 | 0.31 | - | 6.69 | 6.62 | - |
|  |  |  | 2 | 0.50 | 0.144 | 0.095 | 0.250 | 0.297 | - | 0.50 | 0.59 | - | 6.83 | 6.90 | - |
| HD131511 | 05/03/93 | 0.75 | - | - | 0.259 | 0.200 | 0.348 | 0.317 | - | 0.35 | 0.32 | - | 5.90 | 5.86 | - |
| MS Ser | 07/03/93 | 0.16 | H | 0.82/0.18 | 1.965 | 1.719 | 2.004 | 1.832 | 0.535 | 2.44 | 2.23 | 0.65 | 6.75 | 6.71 | 6.18 |
| $\sigma^2$ CrB | 01/02/88 | 0.77 | H | 0.58 | 0.142 | 0.144 | 0.214 | 0.274 | - | 0.37 | 0.47 | - | 6.84 | 6.94 | - |
|  |  |  | C | 0.42 | 0.166 | 0.148 | 0.286 | 0.292 | - | 0.68 | 0.70 | - | 6.88 | 6.88 | - |
| " | 14/07/89 | 0.54 | T | 0.58/0.42 | 0.299 | 0.264 | 0.488 | 0.495 | - | ... | ... | - | ... | ... | - |
| V772 Her | 26/07/88 | 0.45 | T | 0.79/0.21 | 0.277 | 0.204 | ... | ... | - | ... | ... | - | ... | ... | - |
| " | 27/07/88 | 0.53 | T | 0.79/0.21 | 0.471 | 0.311 | ... | ... | - | ... | ... | - | ... | ... | - |
| " | 27/07/88 | 0.62 | T | 0.79/0.21 | 0.471 | 0.343 | ... | ... | - | ... | ... | - | ... | ... | - |
| " | 27/07/88 | 0.67 | T | 0.79/0.21 | 0.616 | 0.510 | ... | ... | - | ... | ... | - | ... | ... | - |
| " | 27/07/88 | 0.57 | T | 0.79/0.21 | 0.493 | 0.407 | ... | ... | - | ... | ... | - | ... | ... | - |
| " | 28/07/88 | 0.66 | T | 0.79/0.21 | 0.598 | 0.523 | ... | ... | - | ... | ... | - | ... | ... | - |
| " | 28/07/88 | 0.72 | T | 0.79/0.21 | 0.501 | 0.381 | ... | ... | - | ... | ... | - | ... | ... | - |
| " | 29/07/88 | 0.89 | T | 0.79/0.21 | 0.442 | 0.305 | ... | ... | - | ... | ... | - | ... | ... | - |
| " | 30/07/88 | 0.95 | T | 0.79/0.21 | 0.391 | 0.313 | ... | ... | - | ... | ... | - | ... | ... | - |
| " | 30/07/88 | 0.89 | T | 0.79/0.21 | 0.432 | 0.335 | ... | ... | - | ... | ... | - | ... | ... | - |
| " | 14/07/89 | 0.86 | H | 0.79 | 0.203 | 0.133 | 0.392 | 0.419 | - | 0.50 | 0.53 | - | 6.74 | 6.77 | - |
|  |  |  | C | 0.21 | 0.075 | 0.069 | 0.116 | 0.158 | - | 0.55 | 0.75 | - | 6.67 | 6.81 | - |
| " | 17/07/89 | 0.28 | H | 0.79 | 0.245 | 0.128 | 0.325 | 0.341 | - | 0.41 | 0.43 | - | 6.66 | 6.98 | - |
|  |  |  | C | 0.21 | 0.092 | 0.092 | 0.181 | 0.208 | - | 0.86 | 0.99 | - | 6.86 | 6.93 | - |

**Table 1.** Continue

| Name | Date | $\varphi$ | E | $S_H/S_C$ | Reconstruction | | Spectral subtraction | | | Corrected EW | | | Absolute flux | | |
|---|---|---|---|---|---|---|---|---|---|---|---|---|---|---|---|
| | | | | | EW (K) | EW (H) | EW (K) | EW (H) | EW (H$\epsilon$) | EW (K) | EW (H) | EW (H$\epsilon$) | log F (K) | log F (H) | log F (H$\epsilon$) |
| V815 Her | 26/07/88 | 0.68 | H | - | 0.673 | 0.506 | 0.744 | 0.580 | 0.167 | 0.74 | 0.58 | 0.17 | 6.80 | 6.69 | 6.15 |
| " | 27/07/88 | 0.71 | H | - | 0.581 | 0.509 | 0.738 | 0.609 | 0.249 | 0.74 | 0.61 | 0.25 | 6.80 | 6.71 | 6.33 |
| " | 27/07/88 | 0.22 | H | - | 0.724 | 0.535 | 0.878 | 0.640 | 0.323 | 0.88 | 0.64 | 0.32 | 6.87 | 6.74 | 6.44 |
| " | 28/07/88 | 0.26 | H | - | 0.672 | 0.486 | 0.743 | 0.626 | 0.189 | 0.74 | 0.63 | 0.19 | 6.80 | 6.73 | 6.21 |
| " | 29/07/88 | 0.35 | H | - | 0.658 | 0.476 | 0.861 | 0.578 | 0.194 | 0.86 | 0.58 | 0.19 | 6.86 | 6.69 | 6.22 |
| " | 30/07/88 | 0.38 | H | - | 0.626 | 0.377 | 0.849 | 0.619 | 0.254 | 0.85 | 0.62 | 0.25 | 6.86 | 6.72 | 6.33 |
| BY Dra | 26/07/88 | 0.22 | H | 0.70 | 2.715 | 2.406 | 3.069 | 2.678 | 0.648 | 4.38 | 3.83 | 0.93 | 6.52 | 6.46 | 5.85 |
| | | | C | 0.30 | 1.726 | 1.519 | 2.041 | 1.805 | 0.579 | 6.80 | 6.02 | 1.93 | 6.03 | 5.98 | 5.49 |
| " | 27/07/88 | 0.23 | H | 0.70 | 2.895 | 2.174 | | | | | | | | | |
| | | | C | 0.30 | 1.666 | 1.492 | | | | | | | | | |
| " | 27/07/88 | 0.71 | H | 0.70 | 2.444 | 2.051 | | | | | | | | | |
| | | | C | 0.30 | 1.071 | 0.825 | | | | | | | | | |
| " | 28/07/88 | 0.73 | H | 0.70 | 2.461 | 2.294 | 2.765 | 2.618 | 0.646 | 3.95 | 3.74 | 0.93 | 6.48 | 6.45 | 5.85 |
| | | | C | 0.30 | 1.337 | 0.910 | 1.413 | 1.078 | 0.266 | 4.71 | 3.59 | 0.89 | 5.87 | 5.76 | 5.15 |
| " | 29/07/88 | 0.88 | T | 0.70/0.30 | 3.609 | 3.464 | 3.951 | 3.650 | 1.296 | | | | | | |
| " | 30/07/88 | 0.89 | T | 0.70/0.30 | 3.774 | 3.449 | | | | | | | | | |
| V775 Her | 27/07/88 | 0.04 | H | - | 1.341 | 1.061 | 1.439 | 1.157 | 0.627 | 1.44 | 1.16 | 0.63 | 6.35 | 6.26 | 5.99 |
| " | 27/07/88 | 0.06 | H | - | 1.450 | 1.231 | 1.552 | 1.240 | 0.555 | 1.55 | 1.24 | 0.56 | 6.38 | 6.29 | 5.94 |
| " | 27/07/88 | 0.37 | H | - | 1.429 | 1.228 | 1.325 | 1.216 | 0.421 | 1.33 | 1.22 | 0.42 | 6.31 | 6.28 | 5.82 |
| " | 28/07/88 | 0.40 | H | - | 1.445 | 1.346 | 1.374 | 1.277 | 0.479 | 1.37 | 1.28 | 0.48 | 6.33 | 6.30 | 5.87 |
| " | 30/07/88 | 0.07 | H | - | 1.371 | 0.947 | 1.432 | 1.105 | 0.473 | 1.43 | 1.11 | 0.47 | 6.35 | 6.24 | 7.87 |
| V478 Lyr | 27/07/88 | 0.40 | H | - | 0.834 | 0.717 | 0.805 | 0.647 | 0.242 | 0.81 | 0.65 | 0.24 | 6.52 | 6.43 | 6.00 |
| " | 27/07/88 | 0.43 | H | - | 0.676 | 0.617 | 0.710 | 0.594 | 0.135 | 0.71 | 0.59 | 0.14 | 6.47 | 6.39 | 5.75 |
| " | 28/07/88 | 0.82 | H | - | 0.669 | 0.548 | 0.758 | 0.647 | 0.226 | 0.76 | 0.65 | 0.23 | 6.50 | 6.43 | 5.97 |
| " | 28/07/88 | 0.85 | H | - | 0.638 | 0.557 | 0.736 | 0.594 | 0.145 | 0.74 | 0.59 | 0.15 | 6.48 | 6.39 | 5.78 |
| " | 30/07/88 | 0.77 | H | - | 0.569 | 0.522 | 0.629 | 0.506 | 0.120 | 0.63 | 0.51 | 0.12 | 6.42 | 6.32 | 5.70 |
| ER Vul | 28/07/88 | 0.04 | H | 0.54 | 0.255 | 0.235 | ... | ... | - | 0.47 | 0.44 | - | 6.72 | 6.68 | - |
| | | | C | 0.46 | 0.433 | | ... | ... | - | 0.94 | ... | - | 6.90 | ... | - |
| " | 28/07/88 | 0.10 | H | 0.54 | 0.261 | 0.034 | ... | ... | - | 0.48 | 0.06 | - | 6.73 | 5.84 | - |
| | | | C | 0.46 | 0.434 | 0.169 | ... | ... | - | 0.95 | 0.37 | - | 6.91 | 6.49 | - |
| " | 30/07/88 | 0.92 | H | 0.54 | 0.185 | | ... | ... | - | 0.34 | ... | - | 6.58 | ... | - |
| | | | C | 0.46 | 0.111 | | ... | ... | - | 0.24 | ... | - | 6.31 | ... | - |
| " | 30/07/88 | 0.00 | H | 0.54 | 0.203 | | ... | ... | - | 0.38 | ... | - | 6.62 | ... | - |
| | | | C | 0.46 | 0.377 | | ... | ... | - | 0.82 | ... | - | 6.84 | ... | - |
| " | 14/07/89 | 0.87 | H | 0.54 | 0.260 | 0.164 | ... | ... | - | 0.48 | 0.30 | - | 6.73 | 6.52 | - |
| | | | C | 0.46 | 0.151 | 0.124 | ... | ... | - | 0.33 | 0.27 | - | 6.45 | 6.36 | - |
| " | 15/07/89 | 0.24 | H | 0.54 | 0.446 | 0.433 | ... | ... | - | 0.83 | 0.80 | - | 6.96 | 6.95 | - |
| | | | C | 0.46 | 0.477 | 0.064 | ... | ... | - | 1.05 | 0.14 | - | 6.94 | 6.07 | - |
| " | 16/07/89 | 0.71 | H | 0.54 | 0.305 | 0.272 | ... | ... | - | 0.57 | 0.50 | - | 6.79 | 6.74 | - |
| | | | C | 0.46 | 0.425 | 0.304 | ... | ... | - | 0.92 | 0.66 | - | 6.89 | 6.75 | - |
| KZ And | 07/12/89 | 0.33 | H | 0.50 | 0.600 | 0.544 | 0.757 | 0.709 | - | 1.51 | 1.42 | - | 6.54 | 6.51 | - |
| | | | C | 0.50 | 0.568 | 0.609 | 0.738 | 0.845 | 0.329 | 1.48 | 1.69 | 0.66 | 6.53 | 6.59 | 6.18 |
| " | 15/12/92 | 0.39 | H | 0.50 | 0.631 | 0.625 | 0.734 | 0.631 | - | 1.47 | 1.26 | - | 6.53 | 6.46 | - |
| | | | C | 0.50 | 0.605 | 0.574 | 0.716 | 0.695 | 0.308 | 1.43 | 1.39 | 0.62 | 6.52 | 6.51 | 6.15 |
| KT Peg | 15/12/92 | 0.27 | H | 0.90 | 0.184 | 0.141 | 0.243 | 0.192 | - | 0.27 | 0.21 | - | 6.36 | 6.26 | - |
| | | | C | 0.10 | 0.095 | 0.084 | 0.088 | 0.169 | - | 0.88 | 1.69 | - | 5.37 | 5.66 | - |

**Table 2.** Ca II H & K lines measures in the observed and subtracted spectrum (Group 2)

| Name | Date | φ | E | $S_H/S_C$ | Reconstruction | | Spectral subtraction | | | Corrected EW | | | Absolute flux | | |
|---|---|---|---|---|---|---|---|---|---|---|---|---|---|---|---|
| | | | | | EW (K) | EW (H) | EW (K) | EW (H) | EW (Hε) | EW (K) | EW (H) | EW (Hε) | log F (K) | log F (H) | log F (Hε) |
| AR Psc | 20/11/86 | 0.33 | C | - | 0.993 | 0.818 | 1.075 | 0.979 | 0.326 | 1.08 | 0.98 | 0.33 | 6.39 | 6.35 | 5.88 |
| " | 21/11/86 | 0.39 | C | - | 0.966 | 0.886 | 1.037 | 0.934 | 0.179 | 1.04 | 0.93 | 0.18 | 6.38 | 6.33 | 5.62 |
| " | 25/11/86 | 0.67 | C | - | 1.046 | 0.869 | 1.014 | 0.959 | 0.263 | 1.01 | 0.96 | 0.26 | 6.37 | 6.34 | 5.78 |
| LX Per | /06/85 | 0.86 | C | 0.50/0.50 | 0.598 | 0.549 | ... | ... | - | 1.20 | 1.10 | - | 6.55 | 6.52 | - |
| " | /06/85 | 0.99 | C | 0.50/0.50 | 0.740 | 0.816 | ... | ... | - | 1.48 | 1.63 | - | 6.65 | 6.69 | - |
| " | /06/85 | 0.12 | C | 0.50/0.50 | 0.624 | 0.532 | ... | ... | - | 1.25 | 1.06 | - | 6.57 | 6.50 | - |
| " | /06/85 | 0.86 | C | 0.50/0.50 | 0.533 | 0.457 | ... | ... | - | 1.07 | 0.91 | - | 6.50 | 6.44 | - |
| UX Ari | 16/12/93 | 0.92 | C | 0.60/0.40 | 1.522 | 1.332 | 1.609 | 1.475 | 0.292 | 1.77 | 1.62 | 0.32 | 6.72 | 6.69 | 5.98 |
| V711 Tau | 21/11/86 | 0.16 | H | 0.21 | 0.199 | 0.164 | 0.249 | 0.266 | - | 1.19 | 1.27 | - | 6.78 | 6.81 | - |
| | | | C | 0.79 | 1.560 | 1.366 | 1.903 | 1.722 | 0.357 | 2.41 | 2.18 | 0.45 | 6.72 | 6.67 | 5.99 |
| " | 25/11/86 | 0.57 | T | 0.21/0.79 | 1.753 | 1.590 | 2.108 | 1.946 | 0.310 | 2.67 | 2.46 | 0.39 | 6.76 | 6.73 | 5.93 |
| " | 30/01/88 | 0.46 | T | 0.21/0.79 | 1.422 | 1.366 | 1.689 | 1.564 | 0.565 | 2.14 | 1.98 | 0.72 | 6.66 | 6.63 | 6.169 |
| " | 31/01/88 | 0.83 | H | 0.21 | | | 0.164 | 0.164 | - | 0.78 | 0.78 | - | 6.60 | 6.60 | - |
| | | | C | 0.79 | 1.315 | 0.853 | 1.343 | 1.418 | 0.236 | 1.70 | 1.80 | 0.30 | 6.56 | 6.59 | 5.81 |
| EI Eri | 31/01/88 | 0.94 | - | - | 0.706 | 0.506 | 0.666 | 0.840 | - | 0.67 | 0.84 | - | 6.55 | 6.66 | - |
| 54 Cam | 21/11/86 | 0.29 | C | 0.73/0.27 | 0.112 | 0.092 | 0.217 | 0.189 | - | 0.80 | 0.70 | - | 6.61 | 6.55 | - |
| " | 25/11/86 | 0.65 | C | 0.73/0.27 | 0.131 | 0.121 | ... | ... | - | 0.49 | 0.45 | - | 6.39 | 6.35 | - |
| " | 26/11/86 | 0.73 | C | 0.73/0.27 | 0.191 | 0.167 | ... | ... | - | 0.71 | 0.62 | - | 6.55 | 6.49 | - |
| " | 31/01/88 | 0.83 | C | 0.73/0.27 | 0.171 | 0.178 | 0.152 | 0.147 | - | 0.56 | 0.54 | - | 6.45 | 6.44 | - |
| HU Vir | 09/03/93 | 0.71 | - | - | 2.375 | 2.608 | 2.674 | 2.573 | 0.794 | 2.67 | 2.57 | 0.79 | 5.85 | 5.84 | 5.33 |
| HD113816 | 07/03/93 | 0.68 | - | - | 2.815 | 2.461 | 2.892 | 2.650 | - | 2.89 | 2.65 | - | 6.13 | 6.09 | - |
| RS CVn | 28/01/88 | 0.86 | C | 0.80/0.20 | 0.371 | 0.296 | 0.667 | 0.654 | 0.182 | 1.63 | 1.60 | 0.44 | 6.77 | 6.76 | 6.21 |
| " | 01/02/88 | 0.69 | C | 0.80/0.20 | 0.408 | 0.240 | 0.487 | 0.581 | 0.077 | 1.19 | 1.42 | 0.02 | 6.64 | 6.71 | 4.83 |
| HR 5110 | 28/01/88 | 0.19 | C | 0.94/0.06 | 0.059 | 0.057 | ... | ... | - | 0.98 | 0.95 | - | 6.21 | 6.20 | - |
| " | 29/01/88 | 0.58 | C | 0.94/0.06 | 0.062 | 0.048 | ... | ... | - | 1.03 | 0.80 | - | 6.24 | 6.12 | - |
| " | 31/01/88 | 0.34 | C | 0.94/0.06 | 0.031 | 0.031 | ... | ... | - | 0.52 | 0.52 | - | 5.93 | 5.93 | - |
| " | 01/02/88 | 0.72 | C | 0.94/0.06 | 0.068 | 0.060 | ... | ... | - | 1.13 | 1.00 | - | 6.28 | 6.22 | - |
| " | 13/07/89 | 0.02 | C | 0.94/0.06 | 0.069 | 0.073 | ... | ... | - | 1.15 | 1.22 | - | 6.28 | 6.31 | - |
| RV Lib | /06/87 | 0.96 | H | 0.56 | 0.716 | 0.648 | ... | ... | - | | | - | | | - |
| | | | C | 0.44 | 0.132 | 0.169 | ... | ... | - | | | - | | | - |
| SS Boo | /06/87 | 0.43 | C | 0.36/0.64 | 0.586 | 0.506 | ... | ... | ... | 0.98 | 0.84 | - | 6.32 | 6.26 | - |
| " | 26/07/89 | 0.33 | C | 0.36/0.64 | 0.657 | 0.557 | 0.823 | 0.693 | 0.212 | 1.37 | 1.16 | 0.35 | 6.47 | 6.40 | 5.88 |
| RT CrB | /06/87 | 0.73 | H | 0.75 | 0.099 | 0.133 | ... | ... | - | 0.19 | 0.25 | - | 6.00 | 6.13 | - |
| | | | C | 0.25 | 0.222 | 0.304 | ... | ... | - | 0.47 | 0.65 | - | 6.35 | 6.48 | - |
| " | 26/07/88 | 0.80 | H | 0.75 | 0.185 | 0.070 | 0.227 | 0.152 | - | 0.43 | 0.29 | - | 6.36 | 6.19 | - |
| | | | C | 0.25 | 0.415 | 0.261 | 0.374 | 0.259 | - | 0.80 | 0.55 | - | 6.57 | 6.42 | - |
| WW Dra | /06/87 | 0.87 | H | 0.71 | 0.087 | 0.050 | ... | ... | - | 0.12 | 0.07 | - | 5.82 | 5.58 | - |
| | | | C | 0.29 | 0.450 | 0.313 | ... | ... | - | 1.55 | 1.08 | - | 6.64 | 6.48 | - |
| " | 27/07/88 | 0.63 | H | 0.71 | 0.062 | | 0.102 | 0.104 | - | 0.14 | 0.15 | - | 5.89 | 5.90 | - |
| | | | C | 0.29 | 0.399 | 0.326 | 0.662 | 0.462 | - | 2.28 | 1.59 | - | 6.81 | 6.65 | - |
| " | 17/07/89 | 0.31 | H | 0.71 | 0.096 | 0.071 | 0.234 | 0.171 | - | 0.33 | 0.24 | - | 6.25 | 6.11 | - |
| | | | C | 0.29 | 0.484 | 0.513 | 0.688 | 0.654 | 0.183 | 2.37 | 2.26 | 0.63 | 6.82 | 6.80 | 6.25 |

**Table 2.** Continue

| Name | Date | $\varphi$ | E | $S_H/S_C$ | Reconstruction | | Spectral subtraction | | | Corrected EW | | | Absolute flux | | |
|---|---|---|---|---|---|---|---|---|---|---|---|---|---|---|---|
| | | | | | EW (K) | EW (H) | EW (K) | EW (H) | EW (H$\epsilon$) | EW (K) | EW (H) | EW (H$\epsilon$) | log F (K) | log F (H) | log F (H$\epsilon$) |
| HR 6469 | /06/87 | | C | 0.24/0.36 | 0.028 | 0.027 | ... | ... | - | 0.04 | 0.04 | - | 5.00 | 4.99 | - |
| " | 27/07/88 | 0.64 | C | 0.24/0.36 | 0.044 | 0.039 | ... | ... | - | 0.06 | 0.05 | - | 5.20 | 5.15 | - |
| " | 13/07/89 | 0.81 | C | 0.24/0.36 | 0.048 | 0.053 | ... | ... | - | 0.06 | 0.07 | - | 5.24 | 5.28 | - |
| Z Her | 29/07/88 | 0.98 | C | 0.73/0.27 | 0.381 | 0.262 | 0.564 | 0.426 | 0.029 | 2.09 | 1.58 | 0.11 | 6.82 | 6.70 | 5.54 |
| " | 14/07/89 | 0.64 | C | 0.73/0.27 | 0.241 | 0.192 | 0.354 | 0.370 | 0.053 | 1.31 | 1.37 | 0.20 | 6.62 | 6.64 | 5.80 |
| " | 15/07/89 | 0.88 | C | 0.73/0.27 | 0.215 | 0.191 | 0.329 | 0.309 | 0.041 | 1.22 | 1.14 | 0.15 | 6.59 | 6.56 | 5.69 |
| " | 16/07/89 | 0.13 | C | 0.73/0.27 | 0.210 | 0.234 | 0.352 | 0.292 | 0.116 | 1.30 | 1.08 | 0.43 | 6.62 | 6.54 | 6.14 |
| " | 17/07/89 | 0.39 | C | 0.73/0.27 | 0.204 | 0.195 | 0.344 | 0.296 | 0.108 | 1.27 | 1.10 | 0.40 | 6.61 | 6.54 | 6.11 |
| MM Her | 27/07/88 | 0.27 | H | 0.69 | 0.107 | 0.060 | 0.236 | 0.174 | - | 0.34 | 0.25 | - | 6.29 | 6.16 | - |
| | | | C | 0.31 | 0.634 | 0.472 | 0.771 | 0.615 | 0.236 | 2.49 | 1.98 | 0.76 | 6.84 | 6.75 | 6.33 |
| " | 16/07/89 | 0.89 | T | 0.69/0.31 | 0.762 | 0.737 | 0.931 | 0.916 | 0.210 | 3.00 | 2.96 | 0.68 | 6.93 | 6.92 | 6.28 |
| AW Her | 26/07/88 | 0.91 | C | 0.54/0.46 | 1.617 | 1.506 | 2.054 | 1.852 | 0.529 | 3.67 | 3.31 | 0.95 | 6.90 | 6.85 | 6.31 |
| 42 Cap | 27/07/88 | 0.18 | - | - | 0.141 | 0.113 | 0.168 | 0.133 | - | 0.27 | 0.13 | - | 5.96 | 5.85 | - |
| RT Lac | 25/11/86 | 0.87 | H | 0.57 | 1.621 | 0.647 | 0.894 | 1.323 | - | 1.57 | 2.32 | - | 6.76 | 6.93 | - |
| | | | C | 0.43 | 1.041 | 0.508 | 1.329 | 0.884 | - | 3.09 | 2.06 | - | 6.82 | 6.65 | - |
| " | 27/07/88 | 0.97 | T | 0.57/0.43 | 2.319 | 1.857 | 2.560 | 2.186 | 0.395 | | | - | | | |
| " | 30/07/88 | 0.55 | T | 0.57/0.43 | 1.684 | 1.366 | 1.881 | 1.781 | - | | | - | | | |
| " | 18/07/89 | 0.12 | H | 0.57 | 1.380 | 0.941 | 1.705 | 1.662 | - | 2.99 | 2.92 | - | 7.04 | 7.03 | - |
| | | | C | 0.43 | 1.410 | 1.208 | 1.714 | 1.425 | 0.457 | 3.99 | 3.31 | 1.06 | 6.93 | 6.85 | 6.36 |
| AR Lac | 14/07/89 | 0.95 | T | 0.44/0.56 | 0.817 | 0.745 | 1.109 | 1.330 | - | 1.98 | 2.38 | - | 6.74 | 6.82 | - |
| " | 18/07/89 | 0.95 | T | 0.44/0.56 | 0.661 | 0.687 | 0.992 | 1.061 | - | 1.77 | 1.90 | - | 6.70 | 6.73 | - |
| SZ Psc | 21/11/86 | 0.40 | C | 0.71/0.39 | 0.648 | 0.732 | ... | ... | ... | 0.93 | 1.05 | - | 6.30 | 6.35 | - |
| " | 21/11/86 | 0.42 | C | 0.71/0.39 | 0.717 | 0.656 | ... | ... | ... | 1.02 | 0.94 | - | 6.34 | 6.31 | - |
| " | 25/11/86 | 0.42 | C | 0.71/0.39 | 0.790 | 0.719 | ... | ... | ... | 1.13 | 1.03 | - | 6.39 | 6.35 | - |
| " | 30/07/88 | 0.57 | C | 0.71/0.39 | 1.073 | 0.644 | 1.446 | 1.079 | 0.376 | 2.07 | 1.54 | 0.54 | 6.65 | 6.52 | 6.06 |
| " | 16/07/89 | 0.32 | C | 0.71/0.39 | 0.707 | 0.574 | 0.974 | 0.925 | 0.110 | 1.39 | 1.32 | 0.16 | 6.48 | 6.46 | 5.53 |

**Table 3.** Ca II H & K lines measures in the observed and subtracted spectrum (Group 3)

| Name | Date | $\varphi$ | E | $S_H/S_C$ | Reconstruction | | Spectral subtraction | | | Corrected EW | | | Absolute flux | | |
|---|---|---|---|---|---|---|---|---|---|---|---|---|---|---|---|
| | | | | | EW (K) | EW (H) | EW (K) | EW (H) | EW (H$\epsilon$) | EW (K) | EW (H) | EW (H$\epsilon$) | log F (K) | log F (H) | log F (H$\epsilon$) |
| 33 Psc | 25/11/86 | 0.19 | - | - | 0.300 | 0.105 | ... | ... | - | ... | ... | - | ... | ... | - |
| 5 Cet | 15/12/92 | 0.32 | C | - | 0.440 | 0.273 | ... | ... | - | 0.44 | 0.27 | - | 5.68 | 5.47 | - |
| BD Cet | 12/12/92 | 0.90 | C | - | 0.984 | 0.699 | 1.147 | 1.005 | - | 1.15 | 1.01 | - | 6.10 | 6.04 | - |
| $\zeta$ And | 24/10/91 | 0.29 | - | - | 0.715 | 0.568 | 0.945 | 0.916 | - | 0.95 | 0.92 | - | 6.01 | 6.00 | - |
| " | 12/12/92 | 0.69 | - | - | 0.784 | 0.636 | 0.980 | 0.957 | - | 0.98 | 0.96 | - | 6.03 | 6.02 | - |
| " | 19/09/93 | 0.02 | - | - | ... | 0.494 | ... | ... | - | ... | 0.49 | - | ... | 5.73 | - |
| " | 01/10/93 | 0.13 | - | - | ... | 0.455 | ... | ... | - | ... | 0.46 | - | ... | 5.64 | - |
| " | 01/10/93 | 0.18 | - | - | ... | 0.418 | ... | ... | - | ... | 0.42 | - | ... | 5.66 | - |
| " | 03/10/93 | 0.30 | - | - | ... | 0.342 | ... | ... | - | ... | 0.34 | - | ... | 5.57 | - |
| " | 06/10/93 | 0.41 | - | - | ... | 0.395 | ... | ... | - | ... | 0.40 | - | ... | 5.63 | - |
| " | 02/11/93 | 0.98 | - | - | ... | 0.520 | ... | ... | - | ... | 0.52 | - | ... | 5.75 | - |
| $\eta$ And | 12/12/92 | 0.62 | T | 0.50/0.50 | 0.102 | 0.065 | 0.077 | 0.057 | - | 0.08 | 0.06 | - | 5.26 | 5.13 | - |
| AY Cet | 12/12/92 | 0.61 | C | - | 0.589 | 0.525 | 0.711 | 0.657 | - | 0.71 | 0.66 | - | 6.26 | 6.23 | - |
| HD 12545 | 15/12/92 | 0.55 | - | - | 4.534 | 3.897 | 4.874 | 4.308 | 1.243 | 4.87 | 4.31 | 1.24 | 6.85 | 6.79 | 6.25 |
| 6 Tri | 15/12/92 | 0.87 | C | 0.20/0.80 | 0.393 | 0.304 | 0.566 | 0.476 | - | 0.61 | 0.51 | - | 5.94 | 5.87 | - |
| 12 Cam | 21/11/86 | 0.50 | - | - | 1.741 | 1.626 | ... | ... | - | 1.74 | 1.63 | - | 6.28 | 6.25 | - |
| " | 21/11/86 | 0.51 | - | - | 1.618 | 1.599 | 2.184 | 2.080 | - | 2.18 | 2.08 | - | 6.38 | 6.35 | - |
| " | 25/11/86 | 0.52 | - | - | 1.704 | 1.612 | ... | ... | - | 1.70 | 1.61 | - | 6.27 | 6.24 | - |
| " | 25/11/86 | 0.57 | - | - | 1.494 | 1.469 | ... | ... | - | 1.49 | 1.47 | - | 6.21 | 6.20 | - |
| " | 26/11/86 | 0.57 | - | - | 1.489 | 1.439 | ... | ... | - | 1.49 | 1.44 | - | 6.21 | 6.19 | - |
| V1149 Ori | 04/03/93 | 0.19 | C | - | 1.697 | 1.570 | 1.971 | 1.915 | 0.277 | 1.97 | 1.92 | 0.28 | 6.05 | 6.04 | 5.20 |
| " | 04/03/93 | 0.19 | C | - | 1.667 | 1.562 | 1.963 | 1.928 | 0.262 | 1.96 | 1.93 | 0.26 | 6.05 | 6.04 | 5.18 |
| CQ Aur | 25/11/86 | 0.42 | C | 0.27/0.73 | 0.443 | 0.484 | ... | ... | - | 0.61 | 0.66 | - | 6.12 | 6.16 | - |
| $\sigma$ Gem | 28/01/88 | 0.82 | - | - | 1.071 | 0.860 | 1.090 | 1.034 | - | 1.09 | 1.03 | - | 6.07 | 6.05 | - |
| " | 29/01/88 | 0.82 | - | - | 1.085 | 0.904 | 1.113 | 1.051 | - | 1.11 | 1.05 | - | 6.08 | 6.06 | - |
| " | 31/01/88 | 0.92 | - | - | 1.151 | 0.968 | 1.078 | 1.032 | - | 1.08 | 1.03 | - | 6.07 | 6.05 | - |
| " | 24/10/91 | 0.41 | - | - | 1.356 | 1.112 | 1.489 | 1.386 | - | 1.49 | 1.39 | - | 6.21 | 6.18 | - |
| " | 29/09/93 | 0.42 | - | - | .. | 1.098 | ... | ... | - | ... | 1.10 | - | ... | 6.08 | - |
| " | 03/10/93 | 0.62 | - | - | .. | 0.849 | ... | ... | - | ... | 0.85 | - | ... | 5.96 | - |
| " | 05/10/93 | 0.72 | - | - | .. | 0.823 | ... | ... | - | ... | 0.82 | - | ... | 5.95 | - |
| " | 06/10/93 | 0.77 | - | - | .. | 0.705 | ... | ... | - | ... | 0.71 | - | ... | 5.88 | - |
| " | 03/11/93 | 0.20 | - | - | .. | 1.092 | ... | ... | - | ... | 1.09 | - | ... | 6.07 | - |
| RZ Cnc | 31/01/88 | 0.36 | H | 0.80 | 1.302 | 1.286 | 1.427 | 1.323 | 0.195 | 1.78 | 1.65 | 0.24 | 6.29 | 6.25 | 5.42 |
| | | | C | 0.20 | | | 0.302 | 0.497 | - | 1.51 | 2.49 | - | 5.75 | 5.97 | - |
| " | 08/03/93 | 0.44 | H | 0.80/0.20 | 1.619 | 1.499 | 1.880 | 1.734 | 0.340 | 2.35 | 2.17 | 0.43 | 6.41 | 6.37 | 5.66 |
| DM UMa | 07/03/93 | 0.85 | - | - | 2.733 | 2.254 | 3.127 | 2.890 | 0.745 | 3.13 | 2.89 | 0.75 | 6.56 | 6.53 | 5.94 |
| 93 Leo | 29/01/88 | 0.43 | C | 0.46/0.54 | 0.086 | 0.095 | ... | ... | - | 0.16 | 0.18 | - | 5.76 | 5.81 | - |
| " | 31/01/88 | 0.46 | C | 0.46/0.54 | 0.082 | 0.103 | ... | ... | - | 0.15 | 0.19 | 5.74 | 5.74 | 5.84 | - |

**Table 3.** Continue

| Name | Date | $\varphi$ | E | $S_H/S_C$ | Reconstruction | | Spectral subtraction | | | Corrected EW | | | Absolute flux | | |
|---|---|---|---|---|---|---|---|---|---|---|---|---|---|---|---|
| | | | | | EW (K) | EW (H) | EW (K) | EW (H) | EW (H$\epsilon$) | EW (K) | EW (H) | EW (H$\epsilon$) | log F (K) | log F (H) | log F (H$\epsilon$) |
| DK Dra | 26/11/86 | 0.44 | T | 0.50/0.50 | 1.253 | 1.140 | 1.464 | 1.388 | - | 1.46 | 1.39 | - | 5.95 | 5.96 | - |
| " | 29/01/88 | 0.10 | T | 0.50/0.50 | 1.496 | 1.377 | 1.614 | 1.489 | - | 1.61 | 1.49 | - | 6.00 | 5.96 | - |
| " | 31/01/88 | 0.13 | T | 0.50/0.50 | 1.503 | 1.155 | 1.640 | 1.464 | - | 1.64 | 1.46 | - | 6.00 | 5.95 | - |
| " | 31/01/88 | 0.13 | T | 0.50/0.50 | 1.298 | 1.312 | 1.594 | 1.493 | - | 1.59 | 1.49 | - | 5.99 | 5.96 | - |
| " | 07/03/93 | 0.02 | T | 0.50/0.50 | 1.844 | 1.700 | 2.007 | 1.924 | 0.208 | 2.01 | 1.92 | 0.21 | 6.09 | 6.07 | 5.11 |
| " | 09/03/93 | 0.05 | T | 0.50/0.50 | 1.730 | 1.558 | 1.870 | 1.783 | 0.164 | 1.87 | 1.78 | 0.16 | 6.06 | 6.04 | 5.00 |
| 4 UMi | 05/03/93 | 0.86 | - | - | 0.100 | 0.065 | 0.145 | 0.107 | - | 0.15 | 0.11 | - | 4.73 | 4.60 | - |
| GX Lib | 13/07/89 | 0.36 | - | - | 0.900 | 0.761 | 1.047 | 1.116 | - | 1.05 | 1.12 | - | 6.06 | 6.08 | - |
| " | 14/07/89 | 0.44 | - | - | 0.619 | 0.538 | 0.754 | 0.849 | - | 0.75 | 0.85 | - | 5.91 | 5.96 | - |
| " | 05/03/93 | 0.83 | - | - | 0.689 | 0.555 | 0.799 | 0.801 | - | 0.80 | 0.80 | - | 5.94 | 5.94 | - |
| $\epsilon$ UMi | 14/07/89 | 0.43 | C | - | 0.161 | 0.116 | 0.281 | 0.327 | - | 0.28 | 0.33 | - | 5.86 | 5.92 | - |
| V792 Her | 15/07/89 | 0.10 | C | 0.40/0.60 | 1.180 | 0.981 | 1.278 | 1.124 | 0.206 | 1.88 | 1.65 | 0.30 | 6.43 | 6.38 | 5.64 |
| DR Dra | 13/07/89 | 0.67 | C | - | 1.164 | 1.045 | 1.320 | 1.296 | 0.155 | 1.32 | 1.30 | 0.16 | 6.16 | 6.15 | 5.23 |
| | 09/03/93 | 0.09 | C | - | 1.445 | 1.098 | 1.553 | 1.375 | 0.237 | 1.55 | 1.38 | 0.24 | 6.23 | 6.17 | 5.41 |
| o Dra | 16/07/89 | 0.66 | - | - | 0.140 | 0.102 | 0.162 | 0.146 | - | 0.16 | 0.15 | - | 5.46 | 5.42 | - |
| V1762 Cyg | 17/07/89 | 0.45 | - | - | 1.002 | 0.871 | 1.086 | 1.021 | - | 1.09 | 1.02 | - | 6.07 | 6.05 | - |
| V1817 Cyg | 28/07/88 | 0.25 | C | 0.10/0.90 | 0.450 | 0.301 | ... | ... | - | 0.50 | 0.33 | - | 5.64 | 5.47 | - |
| " | 30/07/88 | 0.27 | C | 0.10/0.90 | 0.529 | 0.541 | ... | ... | - | 0.59 | 0.60 | - | 5.71 | 5.72 | - |
| " | 18/07/89 | 0.52 | C | 0.10/0.90 | 0.464 | 0.484 | ... | ... | - | 0.52 | 0.54 | - | 5.66 | 5.67 | - |
| V1764 Cyg | 18/07/89 | 0.03 | C | 0.60/0.40 | 1.086 | 1.113 | 1.619 | 1.410 | - | 4.05 | 3.53 | - | 6.64 | 6.58 | - |
| HK Lac | 14/07/89 | 0.89 | C | - | 1.794 | 1.585 | 2.015 | 1.975 | 0.344 | 2.02 | 1.98 | 0.34 | 6.53 | 6.52 | 5.76 |
| V350 Lac | 17/07/89 | 0.50 | - | - | 1.179 | 1.015 | 1.391 | 1.299 | - | 1.39 | 1.30 | - | 6.09 | 6.06 | - |
| IM Peg | 30/07/88 | 0.45 | - | - | 2.038 | 1.603 | 2.427 | 1.926 | 0.212 | 2.43 | 1.93 | 0.21 | 6.14 | 6.04 | 5.08 |
| " | 14/07/89 | 0.65 | - | - | 1.496 | 1.347 | 1.752 | 1.763 | - | 1.75 | 1.76 | - | 6.00 | 6.00 | - |
| $\lambda$ And | 14/07/89 | 0.56 | - | - | 0.865 | 0.803 | 0.965 | 0.949 | - | 0.97 | 0.95 | - | 6.11 | 6.11 | - |
| " | 24/10/91 | 0.10 | - | - | 1.026 | 0.934 | 1.149 | 1.143 | - | 1.15 | 1.14 | - | 6.19 | 6.19 | - |

**Table 4.** Ca II H & K lines measures in the observed and subtracted spectrum(Single stars or components of visual binaries)

| HD | Name | F(1.0 Å) (K) | (H) | Reconstruction EW (K) | EW (H) | Spectral subtraction EW (K) | EW (H) | EW (Hε) | Absolute flux log F (K) | log F (H) | log F (Hε) |
|---|---|---|---|---|---|---|---|---|---|---|---|
| **F** | | | | | | | | | | | |
| 154417 | HR 6349 | 0.199 | 0.230 | 0.049 | 0.051 | 0.112 | 0.164 | - | 6.15 | 6.31 | - |
| **G** | | | | | | | | | | | |
| 115383 | 59 Vir | 0.267 | 0.306 | 0.099 | 0.100 | 0.175 | 0.198 | - | 6.29 | 6.34 | - |
| 206860 | HN Peg | 0.280 | 0.297 | 0.110 | 0.101 | 0.262 | 0.254 | - | 6.46 | 6.45 | - |
| 98231 | ξ UMa A | 0.188 | 0.204 | 0.029 | 0.031 | 0.046 | 0.059 | - | 5.71 | 5.81 | - |
| 218739 | KZ And A | 0.304 | 0.301 | 0.139 | 0.120 | 0.224 | 0.121 | - | 6.39 | 6.13 | - |
| 146362 | $\sigma^1$ CrB | 0.233 | 0.262 | 0.037 | 0.028 | 0.268 | 0.361 | - | 6.47 | 6.60 | - |
| 20630 | $\kappa^1$ Cet | 0.307 | 0.320 | 0.143 | 0.124 | 0.258 | 0.307 | - | 6.34 | 6.42 | - |
| 131156 A | ξ Boo A | 0.428 | 0.428 | 0.233 | 0.229 | 0.477 | 0.463 | - | 6.49 | 6.48 | - |
| 101501 | 61 UMa | 0.262 | 0.277 | 0.087 | 0.066 | 0.121 | 0.098 | - | 5.90 | 5.81 | - |
| **K** | | | | | | | | | | | |
| 190404 | | 0.141 | 0.164 | 0.312 | - | - | - | - | 6.00 | | - |
| 22049 | ε Eri | 0.520 | 0.522 | 0.344 | 0.305 | 0.412 | 0.390 | - | 5.98 | 5.95 | - |
| 4628 | HR 222 | 0.203 | 0.233 | 0.071 | 0.056 | 0.072 | 0.094 | - | 5.22 | 5.34 | - |
| 16160 | HR 753 | 0.216 | 0.222 | 0.073 | 0.045 | - | - | - | 5.00 | 4.79 | - |
| 219134 | HR8832 | 0.183 | 0.206 | 0.065 | 0.052 | - | - | - | 4.95 | 4.85 | - |
| 115404 | | 0.474 | 0.489 | 0.289 | 0.270 | 0.323 | 0.311 | - | 5.65 | 5.63 | - |
| 127665 | ρ Boo | 0.130 | 0.131 | 0.075 | 0.059 | - | - | - | 4.45 | 4.34 | - |
| 131156 B | ξ Boo B | 1.337 | 1.249 | 1.066 | 0.940 | - | - | 0.105 | 5.91 | 5.85 | 4.90 |
| 201091 | 61 Cyg A | 0.659 | 0.655 | 0.453 | 0.381 | - | - | 0.041 | 5.31 | 5.24 | 4.27 |
| 201092 | 61 Cyg B | 1.074 | 1.002 | 0.825 | 0.691 | - | - | 0.039 | 5.12 | 5.05 | 3.79 |